# Training Large Language Models for Advanced Typosquatting Detection

Jackson Welch, MBA
Email: welchjk@mail.uc.edu

*Abstract --* **Typosquatting is a long-standing cyber threat that exploits human error in typing URLs to deceive users, distribute malware, and conduct phishing attacks. With the proliferation of domain names and new Top-Level Domains (TLDs), typosquatting techniques have grown more sophisticated, posing significant risks to individuals, businesses, and national cybersecurity infrastructure. Traditional detection methods primarily focus on well-known impersonation patterns, leaving gaps in identifying more complex attacks. This study introduces a novel approach leveraging large language models (LLMs) to enhance typosquatting detection. By training an LLM on character-level transformations and pattern-based heuristics rather than domain-specific data, a more adaptable and resilient detection mechanism develops. Experimental results indicate that the Phi-4 14B model outperformed other tested models when properly fine tuned achieving a 98% accuracy rate with only a few thousand training samples. This research highlights the potential of LLMs in cybersecurity applications, specifically in mitigating domain-based deception tactics, and provides insights into optimizing machine learning strategies for threat detection.**

## I. Introduction

Since the early days of the commercial internet, typosquatting has exploited the simplest of human errors, mistyping a URL, to serve as a potent tool for cybercriminals. Initially observed as an opportunistic tactic, typosquatting involves registering domain names that closely match that of reputable brands, thereby redirecting users to counterfeit websites. This has evolved into a sophisticated form of cyberattack used to conduct phishing schemes, distribute malware, and harvest sensitive data. Now with billions of domain names and TLDs in circulation, the scale and impact of typosquatting have grown exponentially. This poses significant risks to individuals, businesses, and national cybersecurity infrastructure. This whitepaper explores how emerging large language model (LLM) techniques can enhance the detection of typosquatting attempts, ultimately fortifying defenses against one of the internet's most enduring cyber threats.

Cybercriminals employ various domain squatting techniques to deceive users and bypass traditional security measures. These methods include but not limited to:

**Character Substitution:** These attacks swap similar looking characters like replacing "o" with "0" in **go0gle[.]com** to trick users into believing they are visiting the legitimate site.

**Omission or Addition:** This method involves removing or adding a character, creating domains such as **gogle[.]com** or **gooogle[.]com**, which often go unnoticed by unsuspecting users.

**Homoglyph Attacks:** By using visually similar character combinations (e.g., "rn" for "m"), attackers create domains like **rnicrosoft[.]com** that closely resemble the original.

**Common Misspellings:** Attackers exploit frequent typos, resulting in domains such as **facbook[.]com**, banking on common human error to divert traffic.

**TLD Manipulation:** Instead of the typical ".com" ending, an attacker might use a different TLD—like **paypal[.]co** to deceive users who overlook the subtle difference.

**Phonetic Similarity:** Domains like **nute1ix[.]com** mimic the sound of trusted sites such as **netflix[.]com**, relying on auditory similarity to mislead visitors.

**Deceptive Additions:** By appending words like "support" or "login" to a domain (e.g., **apple-support[.]com**), attackers create a false impression of legitimacy and security [1].

**Punycode:** A standardized encoding method that converts Unicode characters, including those from non-Latin scripts, into an ASCII-compatible format for use in domain names, enabling internationalized domain names (IDNs) like **müller[.]de** to be represented as **xn--mller-kva[.]de**. Threat actors exploit Punycode for typosquatting by registering domain names with visually similar but different Unicode characters, such as replacing the Latin 'a' with the Cyrillic 'a,' tricking users into visiting malicious websites that appear identical to legitimate ones [2].

Significant progress has been made by registrars, legal frameworks, and cybersecurity firms to mitigate these attacks. Modern web browsers have also increased defenses against typosquatting. Today's browsers employ several strategies to alert users to suspicious or potentially fraudulent sites. For instance, browsers such as Google Chrome and Mozilla Firefox now display domain names in Punycode when non-ASCII characters are detected, helping users recognize when a homograph attack might be at play. Browsers also incorporate robust phishing detection systems that analyze web page content and domain reputations in real time, often displaying warnings when a site appears suspect or the security certificate seems mismatched. These built-in features, alongside continuous updates and improved safe browsing protocols, significantly enhance the user's ability to spot and avoid typosquatting attacks [3]. While significant challenges still persist, research has advanced the understanding of fraudulent domain practices. Current detection systems tend to concentrate on established impersonation patterns tied to prominent brands, leaving them less effective against innovative tactics or those targeting lesser-known entities. Techniques that manipulate lower levels of the Domain Name System (DNS), for example, often evade these conventional methods, while the continuously evolving nature of these attacks further strains current defenses. Bridging these gaps is essential for developing a more resilient and comprehensive approach to safeguarding online identity. This, in turn, would ultimately enhance overall cybersecurity for both individuals and organizations. In response, this study introduces a technique of training an LLM on thousands of common and not so common domains against their typosquated counterparts to learn the patterns, rather than the domains themselves, to conduct analysis. This innovation not only offers a more adaptable countermeasure against the dynamic threat landscape but also exemplifies how LLMs can revolutionize the processing of extensive security logs to mitigate online deception.

## II. Retrieval-Automation Generation (RAG) vs Fine Tune Training

Training a dedicated LLM specifically for typosquatting detection has several advantages over a retrieval-augmented generation (RAG) approach. LLMs, trained on specialized data, develop a foundational understanding of the subtle character-level changes with context-specific patterns that differentiate malicious domains from legitimate ones, enabling accurate identification of typosquatting attempts. In contrast, RAG, which combines a retriever for sourcing external documents with a generative model, can be limited by the quality and coverage of its external knowledge base. While RAG can leverage more up-to-date data and a broader context, it may not always capture the evolving and nuanced nature of typosquatting techniques. The reliance on an external knowledge base can introduce vulnerabilities, as it may not be comprehensive or current enough to detect the latest typosquatting strategies. Below, it is discernible that models with fewer parameters struggled to consistently follow the prompt instructions, rendering a RAG-based approach impractical. By training an LLM directly on targeted data, the model inherently learns to detect these subtle variations and can rapidly adapt to new adversarial strategies without needing to consult external repositories, resulting in lower latency and enhanced detection accuracy. Furthermore, a dedicated LLM can be continuously fine-tuned on new data as typosquatting techniques evolve, ensuring that the model remains effective against emerging threats.

### III. Training Data

The initial phase of any machine learning initiative is to establish an accurate representation of the ground truth data. The website **haveibeensquatted[.]com** provided known examples of typosquatted domains paired with their legitimate counterparts [4]. For example, the alteration of **facebook[.]com** to **faceb0ok[.]com** illustrates the underlying concept of swapping an "o" with an "0". This methodology was applied to hundreds of domains, including those associated with companies such as Accenture, Amazon, Cisco, IHG, Tesla, Trendmicro, and Walmart, yielding approximately 500+ real-world domain examples. However, this dataset alone was insufficient for developing a comprehensive model. To address this limitation, a process known as self-distillation through synthetic data generation was employed. This approach leverages a LLM such as ChatGPT O1 or O3-mini, with many more parameters, to generate synthetic training data based on a subset of authentic data points. The larger, more capable "teacher" model produces high-quality datasets that capture specific patterns and domain-specific knowledge, enabling smaller "student" models to internalize nuanced behaviors without the computational burden of training on massive datasets [5]. This methodology accelerates model development, reduces training costs, and allows for the deployment of specialized models optimized for specific applications. The list of true typosquatting domains was provided to ChatGPT O3-mini, along with the prompt: "*Expand on this list and generate many more domain pairings for typosquatting methods.*" The results closely resembled the ground truth dataset, producing examples such as **facebookk[.]com** and **netf1ix[.]com**. To further expand the dataset, a list of companies from the Standard and Poor's (S&P 500) was incorporated, with a similar prompt strategy applied, resulting in an augmented dataset comprising approximately 2,300 verified domain examples for the training set. The validation set was generated using a comparable approach but focused on domains associated with national security, banking, and defense sectors, which are particularly susceptible to spoofing or disguising. By combining authentic domain examples with synthetically generated data via self-distillation, this approach significantly enhances the robustness of the dataset, expedites model development, optimizes resource utilization, and enables the creation of specialized models capable of addressing domain-specific challenges in typosquatting detection.

### IV. Model Selection and Training Setup

Fine-tuning an LLM can require hundreds of graphic processing unit (GPU)-intensive hours. However, to combat this, the Python library Unsloth was leveraged. Unsloth is a lightweight library designed to efficiently fine-tune LLMs, optimizing performance even in resource-constrained environments. It is easily integrated into Google Collab and Linux systems [4]. Many different models were tested such as Meta LLaMA3.2 1B, 3B and LLaMA 3.1 8B, along with DeepSeek R1 distilled 8B. However the best performance was achieved with the Phi-4 14B parameter model. Due to resource limitations, models with more parameters could not be tested. The use of a single RTX 3090 GPU with 24GB of RAM was leveraged. The next step was crafting a system prompt for training. A system prompt sets the initial context, tone, and behavior guidelines for a large language model, shaping how it interprets user input and generates responses. It effectively guides the LLM's role, style, and focus throughout the interaction. ChatGPT was leveraged in a combination of trial and error approaches with the following system prompt:

*"You are an advanced cybersecurity analyst specializing in detecting typosquatting attacks. Your task is to analyze a given domain and determine if it is designed to mislead users by imitating a well-known, legitimate domain. Perform the following detailed analysis:*

*1. Character-Level Comparison*

*Edit Distance Metrics*
*Employ Levenshtein or Damerau-Levenshtein*

*distances to quantify domain differences.*
*Measure and interpret the number and type of modifications, including insertions, deletions, substitutions, or character transpositions.*

### Substitution Analysis
*Identify and evaluate visually or phonetically similar character replacements, such as:*

- *Letter 'o' replaced by number '0'*
- *Letter 'l' replaced by number '1'*

### Transposition and Omission Detection
*Examine domains for swapped characters or missing/extra characters that deviate from expected legitimate domain patterns.*

### 2. Pattern Recognition and Heuristic Analysis

### Common Typo Patterns
*Detect and classify domains based on prevalent typosquatting techniques, including:*

- ***Character Substitution:*** *e.g., go0gle.com vs. google.com*
- ***Omission or Addition:*** *e.g., gogle.com, gooogle.com*
- ***Homoglyph Attacks:*** *e.g., rnicrosoft.com vs. microsoft.com*
- ***Common Misspellings:*** *e.g., facbook.com vs. facebook.com*
- ***TLD Manipulation:*** *e.g., paypal.co vs. paypal.com*
- ***Phonetic Similarity:*** *e.g., nute1ix.com vs. netflix.com*
- ***Deceptive Additions:*** *misleading terms such as support, login, secure, update, verification, helpdesk*
  - *Examples: apple-support[.]com, bankofamerica-login[.]com*

### Visual and Phonetic Similarity Assessment
*Evaluate domains for visual or phonetic changes designed to cause confusion or misunderstanding.*

### Contextual Domain Structure
*Assess the use of deceptive prefixes, suffixes, and domain extensions that may indicate malicious intent.*

### Final Verdict

- *Return* True *if the domain is identified as a typosquat.*
- *Return False if no intentional deceptive practices are detected.*
- *No additional explanations or details beyond the binary verdict are required."*

Once a system prompt has been created, the fine tuning model parameters can be adjusted. Unsloth makes this exceptionally easy with just a single block of code and a few parameters.

***WARMUP_STEPS*** *= 20* gradually increases the learning rate over the first 100 steps to stabilize training and prevent sudden weight updates that could cause divergence, especially when working with smaller datasets.

***NUM_TRAIN_EPOCHS*** *= 1* ensures the model makes a single full pass over the dataset, which is useful for quick fine-tuning runs, avoiding overfitting on large datasets, and enabling faster experimentation, since there are only a few thousand data points.

***MAX_STEPS*** *= 80* limits training to 80 optimization steps, which helps control overfitting, allows for early stopping, and ensures efficient resource usage, especially for smaller datasets. In testing it was discovered the most optimal performance reached at around 80 samples as overfitting occurred past this point.

***LEARNING_RATE*** *= 1E-5* provides a stable, low learning rate that fine-tunes our pre-trained model without drastically altering their weights. This was set low in this case as the model at times would overfit.

The validation dataset consisted of over 400 domains from various companies with 150 of them being valid domains from popular brands like WhatsApp, Chevron and ESET. The typosquat variants were generated using a combination of synthetically generated data and real valid domains. The goal was to create a unique dataset across various industries with a focus on sectors like energy and banking as those tend to be likely targets for typosquatting but would not appear in our training dataset.

## V. Evaluation

To assess the effectiveness of various models in detecting typosquat domains, extensive training was conducted on multiple open-source models, including variants of LLaMA 3.1, 3.2 and DeepSeek R1 8B distilled. Each model was evaluated on its accuracy in distinguishing typosquat domains from legitimate ones as well as its inference speed. The **Phi-4 14B** model consistently outperformed the others, achieving an impressive **98%** accuracy after fine tuning. In contrast, even though the DeepSeek R1 distilled model demonstrated promising reasoning capabilities, it suffered from slower inference times and lower accuracy. It often overthought and would not follow the prompt; usually outputting a nonsensical analysis. Notably, the non-fine-tuned versions of the LLaMA models (1B and 3B) exhibited significant shortcomings also. They both either failed to follow the prompt or outright refused to engage with the task, often flagging the activity as illegal. This behavior underscores the limitations of deploying models without targeted fine tuning, especially in scenarios that demand strict adherence to domain-specific instructions. This highlights the power of Unsloth that with just a few training samples LLaMA 3.2 1B model improved dramatically from a mere 2% accuracy to 92% after fine-tuning. The older LLaMA 3.1 8B variant performed worse compared to the smaller yet newer LLaMA 3.2 3B model, emphasizing advances in newer model architectures. The table below summarizes the performance of each tested model:

| Model Name | Not Fine Tuned | Fine Tuned | Time (seconds) |
|---|---|---|---|
| LLaMA 3.1 8B DeepSeek R1 | 0% | 81% | 3645 |
| LLaMA 3.2 1B | 2% | 92% | 127 |
| LLaMA 3.2 3B | 77% | 94% | 143 |
| LLaMA 3.1 8B | 66% | 94% | 145 |
| Phi-4 14B | 91% | 98% | 167 |

Phi-4 14B model's superior performance is likely attributed to its higher number of parameters, allowing for greater contextual comprehension and pattern recognition. A closer examination of the model's fine turned misclassifications reveals interesting patterns in its decision-making process. One notable false positive occurred with the domain **duke-energy[.]com**, which is the legitimate domain for Duke Energy but was flagged as a likely typosquating attempt. This indicates that while the model effectively identifies deceptive patterns, it may occasionally mistake certain valid domains as fraudulent due to their structural similarities to common typosquatting techniques. These cases highlight the challenge of differentiating between legitimate corporate branding choices and domain names that exhibit characteristics typically associated with phishing or impersonation attempts. The model did however successfully classify domains such as **ihg-hotels[.]com** and **dellsupport[.]com** as typosquating attempts. These results demonstrate the model's ability to detect common deception strategies, including deceptive additions and slight character modifications. The performance disparity between fine-tuned and non-fine-tuned models underscores the necessity of targeted training to ensure that models can reliably execute specialized cybersecurity tasks. The results also highlight the limitations of smaller models in following complex heuristics, as demonstrated by the underwhelming performance of the non-fine-tuned LLaMA 3.2 1B and 3B models. Without additional training, these models either performed poorly in classification or outright refused to generate outputs, making them unsuitable for this specific cybersecurity use case.

## VI. Future Work

Future work could potentially focus on broadening and refining the dataset to encompass a wider variety of industries and emerging typosquatting tactics. This would allow the model to adapt to a more diverse threat landscape. It may also be beneficial to enrich the model with additional cybersecurity information by incorporating more varied threat intelligence and real-world attack patterns to further enhance its detection capabilities. Investigating occasional false positives, such as the misclassification of legitimate domains, could help guide further fine-tuning of both the model and system prompts, potentially improving precision. There is also the possibility of testing with larger models to explore whether increased parameter counts might deliver even greater accuracy and robustness. Additionally, integrating this LLM based detection mechanism into live, real-time cybersecurity frameworks and considering hybrid approaches that blend retrieval-augmented strategies with fine-tuning might offer promising avenues for bolstering defenses against evolving adversarial techniques.

## VII. Conclusion

This research demonstrates the potential of fine-tuned LLMs in proactively detecting typosquatting attacks with high accuracy using a relatively small training dataset. By shifting the focus from static domain lists to pattern-based heuristics, the solution developed a more adaptive and resilient detection approach that can generalize across different typosquatting techniques. While traditional browser security measures and phishing detection systems offer reactive defenses, LLM-driven detection provides a proactive layer of protection, identifying deceptive domains before users fall victim to them. As cyber threats continue to evolve, leveraging advancements in machine learning and cybersecurity will be crucial in safeguarding individuals, businesses, and national critical infrastructure against domain-based deception.